\documentclass[twocolumn,showpacs,preprintnumbers,amsmath,amssymb]{revtex4}

\usepackage{amsmath,amssymb,ams,bbm,epsfig}
\newcommand{\be}{\begin{equation}}
\newcommand{\ee}{\end{equation}}
\newcommand{\ba}{\begin{array}}
\newcommand{\ea}{\end{array}}
\newcommand{\bqa}{\begin{eqnarray}}
\newcommand{\eqa}{\end{eqnarray}}

\newcommand{\ie}{{\it i.e. }}
\newcommand{\tr}{\mbox{Tr}}

\newcommand{\bra}[1]{\ensuremath{\langle #1 |}}
\newcommand{\ket}[1]{\ensuremath{| #1 \rangle}}
\newcommand{\ovl}[2]{\ensuremath{\langle #1 | #2 \rangle}}

\begin{document}

\title{Concurrence of quasi pure quantum states}

\author{Florian Mintert and Andreas Buchleitner}

\affiliation{
  Max Planck Institute for the Physics of Complex Systems
  N\"othnitzerstr. 38
  01187 Dresden
}

\date{\today}

\begin{abstract}
We derive an analytic approximation for the concurrence of weakly
mixed bipartite quantum states - typical objects in state of the art experiments.
This approximation is shown to be a lower bound of the concurrence of arbitrary
states. 
\end{abstract}


\maketitle

Entangled states constitute one of the most fundamental differences
between quantum and classical mechanics.
The objects of prime interest in most experiments are {\em pure} entangled states.
Therefore, experimental setups are designed such that unavoidable environment
coupling is minimised, in order to keep the system state as pure as
possible.
On the theoretical side, most efforts 
are concerned with
arbitrary mixed states -- so far with limited success.
Hardly any entanglement measure can be calculated for systems larger
than bipartite two-level systems --
the smallest ones of interest.

It is astonishing that theory did not proceed more
closely along the lines indicated by experiments,
although approximations inspired by experimental facts
are frequently -- and succesfully -- used in other branches of physics.
In this letter we present an estimate for the entanglement of
almost pure states --
an approach naturally emerging from experimental reality,
where the evolution into a mixed state - due to
non-unitary dynamics --
occurs on a time scale much larger than any other relevant experimental
time scale \cite{blatt03,wein04,wine03}.
Therefore, notwithstanding  the unavoidable evolution of an initially
pure state into a mixed one,
the system state remains {\it quasi pure} during the period of
interest, \ie its density matrix has one single eigenvalue $\mu_1$ that is
much larger than all the other ones.
Under this condition, all terms proportional to integer powers of $\mu_i$,
$i>1$, are small,
and higher order terms can be safely neglected in our subsequent treatment of
concurrence, which is one of the generally accepted entanglement indicators.

The concurrence $c$ of a pure, bipartite quantum state
$\ket{\Psi}\in {\cal H}_1\otimes{\cal H}_2$
can be defined \cite{aud01} as 
\be
c(\Psi)=\sqrt{\tr(\ket{\Psi}\bra{\Psi})^2-\tr_1\varrho_1^2-
\tr_2\varrho_2^2+(\tr\ket{\Psi}\bra{\Psi})^2}\ ,
\label{concdef}
\ee
with the reduced density matrices
$\varrho_1=\tr_2\ket{\Psi}\bra{\Psi}$ and
$\varrho_2=\tr_1\ket{\Psi}\bra{\Psi}$;
the trace over both subsystems is denoted by $\tr$.
An equivalent and 
widely used expression reads
$c(\Psi)=\sqrt{2(\ovl{\Psi}{\Psi}^2-\tr\varrho_r^2)}$,
where $\varrho_r$ is either one of the reduced density matrices.
Though, as we will see in the sequel, the former definition
will turn out to be advantageous in our following generalisation for mixed
states $\varrho$, where concurrence is defined through  
the convex roof 
\be
c(\varrho)=\inf_{\{p_i,\ket{\Psi_i}\}} \sum_i p_i c(\Psi_i), \hspace{.3cm}
\varrho=\sum_i p_i \ket{\Psi_i}\bra{\Psi_i}, \hspace{.3cm}
p_i>0\ .
\ee
Here, the infimum is to be found among all ensembles $\{p_i,\ket{\Psi_i}\}$
representing $\varrho$.
If we start from a given decomposition ({\it e.g.}, the eigensystem
of $\varrho$, $\varrho\ket{\Phi_i}=\mu_i\ket{\Phi_i}$, $i=1,\hdots,n$),
all ensembles can be parametrised through a left-unitary transformation
$V\in{\mathbbm C}^{N\times n}$ \cite{schroed}, where possibly $N\ge n$:
\be
\sqrt{p_i} \ket{\Psi_i}=\sum_{j=1}^N V_{ij} \sqrt{\mu_j}\ket{\Phi_j}\ ,
\hspace{.5cm}
\sum_{i=1}^N V^\dagger_{ki}V_{ij}=\delta_{j,k}\ .
\label{allens}\ee
The concurrence of a mixed state $\varrho$ can therefore be expressed
as \cite{bad02,flo04}

\be
c(\varrho) =\inf_V\sum_i
\left(\left[ V\otimes V {\cal A} \hspace{.1cm}
V^\dagger\otimes V^\dagger\right]_{ii}^{ii}\right)^\frac12 ,
\label{conc}
\ee
where the tensor ${\cal A}$  is defined as
\bqa
{\cal A}_{jk}^{lm}&=&\sqrt{\mu_j\mu_k\mu_l\mu_m}\times
\nonumber\\
&\Big[&
\tr\left(\ket{\Phi_j}\ovl{\Phi_l}{\Phi_k}{\Phi_m}\right)- \nonumber\\
&&\tr_1\left(
\tr_2\left(\ket{\Phi_j}\bra{\Phi_l}\right)
\tr_2\left(\ket{\Phi_k}\bra{\Phi_m}\right)\right)- \nonumber\\
&&\tr_2\left(
\tr_1\left(\ket{\Phi_j}\bra{\Phi_l}\right)
\tr_1\left(\ket{\Phi_k}\bra{\Phi_m}\right)\right)+ \nonumber\\
&&\tr\left(\ket{\Phi_j}\bra{\Phi_l}\right)
\tr\left(\ket{\Phi_k}\bra{\Phi_m}\right)\Big]\ .
\eqa
$\cal A$ is a positive, hermitian operator satisfying
${\cal A}_{jk}^{lm} = {\cal A}_{kj}^{ml} = {\cal A}_{kj}^{lm}$.
The latter symmetry is inherited from the 
symmetric definition of concurrence in
eq. (\ref{concdef}),
and is crucial for our further analysis.
\begin{figure*}[t]
\epsfig{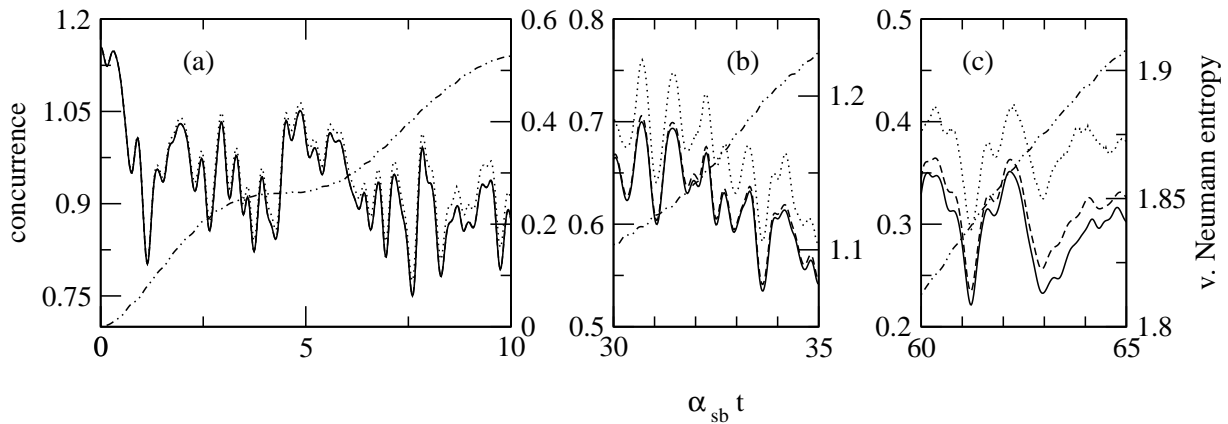}
\caption{Concurrence (left $y$-axis labels) 
in qpa (solid line), together with the optimized lower (dashed
  line) \cite{flo04} and
  upper (dotted line) bounds \cite{flodiss}
  on concurrence for a $3\times 5$ state $\varrho$
  that evolves according to a random, non-unitary time evolution.
  Different degrees of mixing are 
  indicated by the
  von Neumann entropy (double-dotted dashed line; right y-axis labels).
  The gap between the qpa and the upper bound increases from (a) to
  (c), with 
  increasing mixing of $\varrho$.
  Though, even for significantly mixed states this difference does not exceed 
  $0.06$, and the qpa thus provides a reliable estimate of concurrence 
in a wide parameter range. 
}
\label{fig1}
\end{figure*}
In particular, it implies that $\cal A$ can always be expressed in terms of
complex symmetric matrices $T^\alpha$ as
${\cal A}_{jk}^{lm}=\sum_\alpha T^\alpha_{jk}(T^\alpha_{lm})^\ast$ \cite{flo04}.
Whereas the concurrence of a pure state is characterized by a single matrix
$\tau:=T^1$, with only one non-vanishing element,
a mixed state in general requires more than one matrix.
The crucial idea 
underlying our approximation is now that the concurrence of a quasi
pure state can still be well described by a single matrix $\tau$, though with
more than just one non-vanishing element.

This becomes clear 
when 
we 
remind ourselves of the obvious proportionality
${\cal A}_{jk}^{lm}\sim\sqrt{\mu_j\mu_k\mu_l\mu_m}$.
For quasi pure states with $\mu_1\gg\mu_i$, $i>1$,
this relation induces a natural
order in terms of the small eigenvalues $\mu_i$, $i>1$.
The leading term ${\cal A}_{11}^{11}\sim\mu_1^2$ (lowest
order in the $\mu_i$) is sufficient to
characterize the concurrence of 
the pure state $\ket{\Phi_1}$.
For mixed states, however, there are also terms of first order,
alike ${\cal A}_{j1}^{11}\sim\sqrt{\mu_j}$, $j>1$, of
second order, such as ${\cal A}_{jk}^{11}\sim\sqrt{\mu_j\mu_k}$, $j,k>1$,
and of third and fourth order.
The approximation
\be
{\cal A}_{jk}^{lm}\simeq \tau_{jk}\tau_{lm}^\ast\ ,
\hspace{.3cm}\mbox{with}\hspace{.3cm}
\tau_{jk}=\frac{{\cal A}_{jk}^{11}}{\sqrt{{\cal A}_{11}^{11}}}\ ,
\label{Tdef}
\ee
is exact up to first order, and even correctly represents second order
elements of 
type ${\cal {\cal A}}_{jk}^{11}$.
Importantly, this quasi pure approximation (qpa) simplifies eq. (\ref{conc})
significantly:
\be
c(\varrho)\simeq c_{qp}(\varrho)=\inf_V\sum_i
\left|\left[V\tau V^T\right]_{ii}\right|\ ,
\label{cqp}
\ee
and a closed expression for the right hand side of this equation is
known \cite{Wot98, Uhl00}.
It can be given 
in terms of the singular values $\lambda_i$ of $T$,
this is the square roots of the eigenvalues of the positive
hermitian matrix $\tau\tau^\dagger$,
\be
c_{qp}(\varrho)=
\mbox{max}(\lambda_1-\sum_{i>1}\lambda_i,\ 0)\ , 
\label{qpa}
\ee
with the $\lambda_i$ labeled in decreasing order.

Note that, in eq. (\ref{Tdef}), we 
implicitly assumed
that
${\cal A}_{11}^{11}$ does not vanish, or, equivalently, that $\ket{\Phi_1}$
is not separable.
This does not
limit, however, the range of applicability of our
approximation.
If the dominant contribution $\ket{\Phi_1}$ to a given mixed state $\varrho$
is separable, $\varrho$ will typically be separable anyway.

A major advantage of our present estimate is that -- with respect to other
methods \cite{tsw03,flo04} -- it even 
further reduces the
computational resources for the evaluation of the degree of entanglement of a
given state:
In \cite{flo04}, it was necessary to diagonalize a matrix quadratically larger than
$\varrho$, eventually followed by an optimization procedure. Here, we
need not optimize, and only have to diagonalize a matrix of
the size of the given statistical operator.
Consequently, significant speed-up can be achieved when a large number of
mixed states has to be assessed, such as, e.g., monitoring
the time evolution of entanglement under environment coupling \cite{arrc},
without significant loss of the quality of the estimation.
To illustrate that, 
let us consider a bipartite system of dimension $3\times 5$,
coupled to an environment. 
The total Hamiltonian acting on system and bath reads
$H=\alpha_sH_s\otimes{\mathbbm 1}_b+\alpha_{sb}H_{sb}$,
where $H_s$ acts on the system alone, $H_{sb}$ represents the 
system-bath interaction, and
$\alpha_s$ and $\alpha_{sb}$ determine the strength of system dynamics
and environment coupling, respectively. In our following example, the
matrix elements of $H_s$ and $H_{sb}$ are random numbers, and the coupling
constants are fixed at $\alpha_s=0.2$ and
$\alpha_{sb}=0.02$.
The total state of system and bath, initially prepared in a pure system state
tensored with the maximally mixed bath state, evolves 
under the unitary
time evolution operator
$U=\exp(iHt)$,
and the dynamics of the system state alone is obtained upon tracing
over the environmental degrees of freedom,
what induces decoherence, {\it i.e.} mixing. 
In fig. \ref{fig1} the time evolution of concurrence 
is plotted 
in qpa 
for the above parameter values, 
over three different time
intervals. 
Also optimized lower \cite{flo04} 
and upper \cite{flodiss}
bounds of concurrence are shown. 
The degree of mixing of $\varrho(t)$ is characterized by its von
Neumann entropy. 
\begin{figure}
\epsfig{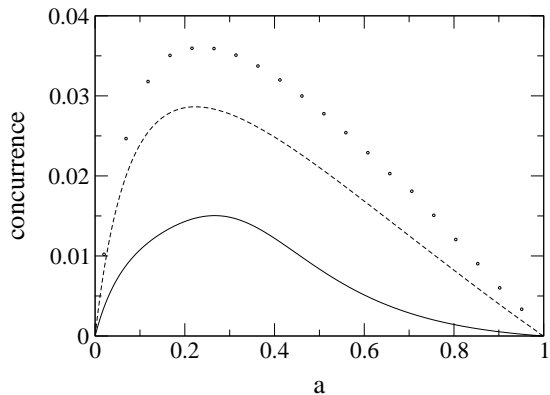}
\caption{
  Concurrence in qpa for the class of $3\times 3$ states defined in 
eq. (\ref{hor}).
  Although $\varrho_a$ exhibits rather large mixing (with von Neumann entropy
approx. equal to $1.3\ldots 1.8$, for $a>0.1$) and positive partial
  transpose, the qpa (solid line) detects $\varrho_a$ as entangled,
  and provides a rather good approximation
  of the actual value of concurrence, which is 
  confined by upper \cite{flodiss}
  (dots) and lower \cite{flo04} (dashed line) bounds.}
\label{fig2}
\end{figure}
While the degree of mixing is rather small 
in fig. \ref{fig1}(a), 
it steadily increases in figs.~\ref{fig1}(b) and \ref{fig1}(c).
Nonetheless, our qpa captures the actual value of the concurrence 
rather precisely over the entire time interval,
even for already significantly mixed states -- i.e. beyond its initially
anticipated range of validity.

Finally, let us note that the quasi pure approximation also provides a lower
bound on 
concurrence, and, in particular, can distinguish entangled states from
separable ones: 
We have shown in \cite{flo04} that any symmetric matrix
$\sum_\alpha z_\alpha T^\alpha$, with $\sum_\alpha |z_\alpha|^2=1$
and
${\cal A}_{jk}^{lm}=\sum_\alpha T^\alpha_{jk}(T^\alpha_{lm})^\ast$,
defines a lower bound of concurrence.
One easily verifies that $\tau$ defined in eq. (\ref{Tdef}) indeed is
precisely of this form, with
$z_\alpha=(T^\alpha_{11})^\ast/\sqrt{\sum_\beta|T^\beta_{11}|^2}$.
Even entangled states with positive partial transpose can be characterised 
with the qpa:
Consider, for example, the class of states 
\be
\varrho_a=\frac{1}{1+8a}\left[
{\renewcommand{\arraystretch}{0.8}\ba{ccccccccc}
a & 0 & 0 & 0 & a & 0 & 0 & 0 & a \\
0 & a & 0 & 0 & 0 & 0 & 0 & 0 & 0 \\
0 & 0 & a & 0 & 0 & 0 & 0 & 0 & 0 \\
0 & 0 & 0 & a & 0 & 0 & 0 & 0 & 0 \\
a & 0 & 0 & 0 & a & 0 & 0 & 0 & a \\
0 & 0 & 0 & 0 & 0 & a & 0 & 0 & 0 \\
0 & 0 & 0 & 0 & 0 & 0 & \beta & 0 & \gamma \\
0 & 0 & 0 & 0 & 0 & 0 & 0 & a & 0 \\
a & 0 & 0 & 0 & a & 0 & \gamma & 0 & \beta
\ea}\right],\ a\in[0,1]\ ,
\label{hor}
\ee
of a $3\times 3$-system, with
$\beta=(1+a)/2$ and
$\gamma=\sqrt{1-a^2}/2$,
which were introduced in \cite{hor97}.
Fig. \ref{fig2} compares the qpa to upper and lower 
bounds \cite{flodiss,flo04} of concurrence, as a function of the parameter $a$.
The qpa 
is indeed positive in the
entire interval, {\it i.e.} the non-separability of $\varrho_a$ is detected 
by purely algebraic means.

Thus, our qpa does not only provide tools for an efficient
estimation of concurrence of states with moderate mixing,
but it is even applicable to general states.
We reckon that the concept of quasi purity which we have exploited here 
may be a remedy also for
various other, so far virtually uncomputable entanglement measures,
and it appears promising to check whether 
such approximations have an 
equally large range of applicability.

We are indebted to Andr\'e Ricardo Ribeiro de Carvalho and 
Marek Ku{\'s} for fruitful discussions,
comments and remarks.
Financial support by VolkswagenStiftung is gratefully acknowledged.

\bibliography{referenzen}

\end{document}